\documentclass[11pt]{article}
\usepackage[utf8]{inputenc}
\usepackage[T1]{fontenc}
\usepackage{lmodern}
\usepackage[a4paper,margin=1in]{geometry}
\usepackage{amsmath,amssymb,amsfonts,bm}
\usepackage{physics}
\usepackage{hyperref}
\usepackage{microtype}
\usepackage{cite}
\usepackage{enumitem}
\usepackage{indentfirst}
\usepackage{tikz}
\usepackage{tikz-feynman}
\tikzfeynmanset{compat=1.1.0}
\setlist[itemize]{leftmargin=1.2em}
\usepackage{authblk}

\begin{document}
\title{From Quantum Tsallis Entropy to Strange Metals}

\date{\today}
\author{
Xian-Hui Ge, \thanks{gexh@shu.edu.cn}
}

\affil[1]{Department of Physics, College of Sciences, Shanghai University, 
99 Shangda Road, Shanghai 200444, China}

\affil[1]{Shanghai Key Laboratory of High-Temperature Superconductivity, 
Shanghai University, Shanghai 200444, China}
\maketitle

\begin{abstract}
We develop a unified framework connecting quantum Tsallis statistics to electronic transport in strongly interacting systems. Starting from Rényi and Tsallis entropies, we construct a quantum Tsallis distribution that reduces to the conventional Fermi--Dirac distribution when $q=1$. For $q$ slightly deviating from unity, the correction term in the occupation function can be mapped to a $q$-deformed Schwarzian action, corresponding to soft reparametrization modes. Coupling these soft modes to electrons via the Fermi Golden Rule yields a modified scattering rate, which reproduces conventional Fermi-liquid behavior at low temperatures and linear-in-temperature resistivity at high temperatures. Using the memory matrix formalism, we analyze magnetotransport, finding a linear-in-field magnetoresistance and a Hall angle consistent with Anderson's two-lifetime scenario. At sufficiently low temperatures, both magnetoresistance and Hall response smoothly recover Fermi-liquid quadratic behaviors. This approach provides a controlled interpolation between Fermi-liquid and non-Fermi-liquid regimes, quantitatively linking $q$-deformation, soft-mode dynamics, and experimentally measurable transport coefficients in strange metals.
\end{abstract}

\hspace{2em} {\emph{Introduction}}-- Landau's Fermi liquid theory has long served as the foundational framework for understanding conventional metals, wherein low-energy excitations are described by long-lived quasiparticles with renormalized effective masses. In this picture, electrical resistivity saturates at both low and high temperatures---due to phonon freezing and the Mott--Ioffe--Regel (MIR) bound, respectively~\cite{Landau1957, PinesNozieres}. However, in a growing class of materials known as \emph{strange metals}, the resistivity remains linear in temperature over a remarkably wide range without saturation, even beyond the MIR limit, indicating the breakdown of the quasiparticle description and the emergence of collective, nonlocal charge-carrying degrees of freedom~\cite{Anderson1990, Varma1997, Emery1995, Lee1992,Phillips2022, Cooper2009}. This non-Fermi liquid behavior in strange metals is typically characterized by linear-in-temperature ($T$-linear) resistivity and linear-in-magnetic-field ($B$-linear) magnetoresistivity, alongside scale-invariant dissipation governed by the Planckian time $\tau_P = \hbar / (k_B T)$~\cite{Anderson1991, Sachdev2011, Zaanen2004, Anderson2013, Bruin2013, Hartnoll2015}.  
Interestingly, the Hall angle in the cuprates exhibits a distinct scaling $\cot \theta_H \sim T^2$, resembling Fermi-liquid-like behavior and therefore standing in sharp contrast to the ubiquitous $T$-linear resistivity of strange metals \cite{Anderson1991,Anderson2013}. Experimental studies — notably the early work by Chien \emph{et al.} — observed such Hall-angle scaling and its sensitivity to impurity scattering, underscoring the inadequacy of single-relaxation-time transport models and pointing instead to multiple, possibly channel-dependent, scattering mechanisms underlying strange-metal phenomenology~\cite{Chien1991}.

\hspace{2em} In gravitational physics, a similarly radical departure from standard local descriptions arises in the resolution of the black hole information loss paradox via the \emph{island prescription}~\cite{Penington2019, Almheiri2019island, Almheiri2020,wormhole9,yang,Almeri,ge25}. In the Euclidean gravitational path integral, two dominant saddles contribute to the entanglement entropy of Hawking radiation: the ``Hawking saddle''---which governs early times and leads to a monotonically increasing entropy---and the ``replica wormhole saddle''---which dominates at late times and produces entropy saturation consistent with unitarity. The von Neumann entanglement entropy is computed using the replica trick,
\begin{equation}
S = \lim_{q \to 1} S_q^{\mathrm{R\acute{e}nyi}} 
= \lim_{q \to 1} \frac{1}{1-q} \log \mathrm{Tr}\,\rho^q,
\label{Rényi}
\end{equation}
where $\rho$ is the reduced density matrix of the Hawking radiation. Geometrically, replica wormholes connect multiple copies of the spacetime, linking the island region with the radiation region in such a way that their mutual entanglement is effectively canceled. This nonlocal connection underlies the unitarity of the full evaporation process.

\hspace{2em} The retrieval of quantum information in gravity thus depends crucially on nonlocal saddle-point contributions. An intriguing question is whether analogous ideas can be applied to strongly correlated condensed matter systems such as strange metals, where transport deviates strongly from Fermi liquid theory and scale-invariant features suggest emergent nonlocal dynamics~\cite{Hartnoll2018, Sachdev2015}. We propose that the \emph{Tsallis entropy}~\cite{Tsallis1988, Tsallis2009}, a nonextensive generalization of Boltzmann--Gibbs entropy, provides a natural framework for such an extension. The Tsallis parameter $q$ encodes deviations from extensivity, modeling systems with long-range correlations and power-law distributions.

\hspace{2em} From this viewpoint, the standard Fermi--Dirac distribution, and the corresponding Fermi liquid behavior, is analogous to the Hawking saddle: a disconnected, local description valid in regimes where quasiparticles dominate. Allowing the replica index $q$ to deviate slightly from unity, $q = 1 + \delta(T)$ with $\delta(T)=T/T_0 < 1$ temperature dependent, yields a Tsallis-deformed fermionic distribution that encodes non-Fermi liquid behavior, reminiscent of the contribution of replica wormholes in semiclassical gravity. Such a deformation can be captured by a Schwarzian effective action~\cite{ Kitaev2015, MaldacenaStanford2016}, which governs the soft mode dynamics of  nearly-$\mathrm{AdS}_2$ geometries. 
 This correspondence suggests a dual statistical picture: the strange metal phase, with its scale-invariant transport and Planckian dissipation, may be understood as the condensed matter analogue of gravitational systems dominated by wormhole saddles. In the following, we develop this idea by deriving the quantum Tsallis distribution, relating it to both Rényi and von Neumann entropies, and showing how its low-energy limit maps to the Schwarzian theory that appears in the strongly interacting quantum matter.

\hspace{2em} {\emph{Quantum Tsallis statistics}}-
For a quantum system described by a density matrix \( \rho \), the Tsallis entropy of order \( q \in \mathbb{R} \) is defined as
\begin{equation}
S_q^{\mathrm{Tsallis}} = \frac{1 - \Tr(\rho^q)}{q - 1}, \qquad (q \neq 1),
\end{equation}
The Tsallis entropy is closely related to the Rényi entropy, defined in equation (\ref{Rényi})
and the two obey the following relation
\begin{equation}
S_q^{\mathrm{Tsallis}} = \frac{1}{q-1} \left[1 - e^{(1 - q) S_q^{\mathrm{Rényi}}} \right].
\end{equation}
This expression shows that the Tsallis entropy provides a power-law expansion of the exponential Rényi entropy. In the limit \( q \to 1 \), both Tsallis and Rényi entropies smoothly reduce to the von Neumann entropy,
\[
S_{\rm vN}=\lim_{q \to 1} S_q^{\mathrm{Tsallis}} = -\Tr(\rho \log \rho).
\]

The Tsallis entropy is non-additive for independent systems, reflecting possible long-range correlations 
\[
S_q^{\mathrm{Tsallis}}(\rho_A\otimes\rho_B)
= S_q^{\mathrm{Tsallis}}(\rho_A)
+ S_q^{\mathrm{Tsallis}}(\rho_B)
+ (1-q)\,S_q^{\mathrm{Tsallis}}(\rho_A)\,S_q^{\mathrm{Tsallis}}(\rho_B).
\]

From the Tsallis and Rényi entropies, we can obtain the corresponding distribution. Using the Langrangian multiplier method with constraints $Tr\rho=1$ and ${\rm Tr}(\rho H)=\langle E\rangle$, we can write down the Langrangian as follows
\begin{equation}
    \mathcal{L}=\frac{1}{1-q}\ln \bigg({\rm Tr} \rho^q\bigg)-\alpha ({\rm Tr} \rho-1)-\beta\bigg({\rm Tr}(\rho H)-\langle E\rangle\bigg).
\end{equation}
We then obtain 
\begin{equation}
   \delta \mathcal{L}=\frac{q}{1-q}\frac{{\rm Tr}(\rho^{q-1}\delta\rho)}{{\rm Tr}(\rho^q)}-\alpha {\rm Tr} \delta \rho-\beta {\rm Tr}(H \delta \rho),
   \end{equation}
   where $\alpha=\mu/T$ and $\beta=1/T$. 
From maximal entropy $\delta \mathcal{L}=0$ and after some manipulation, we find 
\[
\;
\rho \;=\; \frac{1}{Z_q}\,\Big[\,1-(1-q)\beta H\,\Big]_+^{\frac{1}{1-q}}
\;,
\]
 where \([X]_+\) denotes the positive part of the operator 
\([X]\), i.e. the spectral projection onto eigenvalues satisfying \(1-(1-q)\beta \varepsilon_i>0\). This guarantees $\rho \ge 0$ and encodes the Karush–Kuhn–Tucker boundary condition of the constrained maximization \cite{kkt,kkt2}. The normalization \(\Tr\rho=1\) fixes the partition function,
\[
\;
Z_q \;=\; \Tr\!\left(\Big[\,1-(1-q)\beta H\,\Big]_+^{\frac{1}{1-q}}\right)
\;,
\]
which is finite whenever the trace exists; for unbounded \(H\), one either has a high-energy cutoff for \(q<1\) (since \(1-(1-q)\beta\varepsilon\ge0\) implies \(\varepsilon\le 1/[(1-q)\beta]\)) or a power-law tail for \(q>1\) with convergence controlled by the growth of the spectral density. 
For quantum statistics with indistinguishable particles, the same \(q\)-deformation propagates to the grand-canonical occupation numbers once one replaces the ordinary exponential by the \(q\)-exponential while enforcing the exchange symmetry. The generalized quantum occupations can be written compactly as
\[
n_q(\varepsilon)\;=\;
\frac{1}{\big[\,1-(1-q)\beta(\varepsilon-\mu)\,\big]_+^{\frac{1}{q-1}}+a},
\]
where \(\beta=1/T\) and \(\mu\) is the chemical potential; the parameter \(a=+1\) gives the Tsallis–Fermi–Dirac case (Pauli exclusion), \(a=-1\) the Tsallis–Bose–Einstein case (Bose enhancement), and \(a=0\) the Tsallis–Boltzmann classical limit. This expression reduces to the standard Fermi–Dirac/Bose–Einstein forms as \(q\to1\). Using the alternative convention commonly employed in the literature,
the denominator may be written as
$
\big[\,1+(q-1)\beta(\varepsilon-\mu)\,\big]^{\frac{1}{q-1}} + a .
$
  The deformation parameter $q$ organizes the ensemble into two regimes related by the
thermodynamic duality $q \leftrightarrow (2-q)$.  For $q>1$, the distribution exhibits
non-extensive algebraic tails and unbounded energy support, while $q<1$ yields a 
compact-support ensemble determined by the spectral cutoff.  In strange metals we focus
on $q>1$, consistent with  the observed long-tailed 
responses.

\noindent
\hspace{2em} {\emph{q-deformed Schwarzian action-}} We show that the Tsallis-modified fermion distribution can be interpreted as arising from an effective reparametrization theory for the low-energy modes of nearly-$\mathrm{AdS}_2$ gravity. 
In the Tsallis framework, deviations from the Fermi-Dirac (FD) distribution are parameterized by $(q-1)$, which controls the strength of non-extensive effects. 
At $q=1$, the standard FD form is recovered.
For $q\neq 1$, the distribution acquires a correction term $\phi(x)$ that is quadratic in the dimensionless energy variable $x$ and weighted by the thermal factor $n_{\mathrm{FD}}(x)[1-n_{\mathrm{FD}}(x)]$. 
We reinterpret $x$ as the Euclidean frequency operator, $x \leftrightarrow -i\partial_\tau$, and promote $\phi(x)$ to a local functional $\Phi[f](\tau)$ of the time-reparametrization mode $f(\tau)$. 
This functional encodes the same fluctuation structure as $\phi(x)$, but now in the geometric language of boundary diffeomorphisms. 
In the $q\to 1$ limit, $\phi(x)$ vanishes and $\Phi[f](\tau)$ disappears, leaving only the Schwarzian sector; this ensures a smooth interpolation between the Tsallis-deformed and undeformed cases.

\hspace{2em} We start from the expansion of the quantum Tsallis distribution around $q=1$:
\begin{equation}
    n_q(x) = n_{\mathrm{FD}}(x) + (q-1)\,\phi(x) + \cdots, 
    \qquad x = \beta(\epsilon - \mu),
\end{equation}
where
\begin{equation}
    \phi(x) = \frac{x^2}{2}\, n_{\mathrm{FD}}(x)\big[1-n_{\mathrm{FD}}(x)\big].
\end{equation}
Thus, $\phi(x)$ can be interpreted as the variance of the Fermi--Dirac (FD) occupation number, modulated by a quadratic prefactor $x^2$. Since $n_{\mathrm{FD}}(1-n_{\mathrm{FD}})$ is maximized at $x=0$ and decays away from the Fermi surface, this thermal weight confines the correction to states near the Fermi level. In the frequency--time correspondence (taking the Euclidean Matsubara convention), we have $x=\beta \omega$ and $i\partial_\tau \leftrightarrow \omega$. Hence,
\begin{equation}
    x^2 \ \longleftrightarrow\ \beta^2 \omega^2 
    \ \longleftrightarrow\ -\beta^2 \partial_\tau^2 .
\end{equation}
This indicates that the quadratic factor $x^2$ corresponds to the $\omega^2$ contribution in frequency space, which in the time domain is equivalent to a second-order Euclidean time derivative (with an overall sign depending on Fourier conventions). Under the assumption of low-frequency dynamics and a sufficiently smooth response kernel, the frequency-domain correction $\phi(x)$ can be promoted to a local time-domain functional $\Phi[f](\tau)$. In this sense, the $(q-1)$ correction to the distribution can be reinterpreted as a soft-mode contribution involving second-order time derivatives of the reparametrization mode $f(\tau)$. This identification is approximate: for strongly nonlocal frequency responses $|x|\gg 1$, the local replacement by $\partial_\tau^2$ is no longer valid. Motivated by the structure of \(\phi(x)\), we define the local functional
\[
\Phi[f](\tau) = \big(u'(\tau)\big)^2 - 2\, u''(\tau), \qquad u(\tau) = \log f'(\tau),
\]
where \(\left(\frac{f''}{f'}\right)^2 = (u')^2\) and \(\partial_\tau \log f'(\tau) = u'(\tau)\). The quadratic term in \(\phi(x)\) corresponds to \((u')^2\) in the geometric language, while the remaining structure naturally maps to the second derivative term \(u''(\tau)\). The \(q\)-deformed effective action is then schematically written as
\begin{equation}
S_q[f]=S_{sch}+  \Delta S_{q} = -C \int \{f, \tau\} \, + \, (q-1)\,\lambda \int \Phi[f](\tau) \, + \, \mathcal{O}((q-1)^2),
\end{equation}
where \(\{f, \tau\}\) is the Schwarzian derivative and \(\lambda\) is a coupling determined by the microscopic model. For small deviations from identity reparametrization, 
\(f(\tau) = \tau + \varepsilon(\tau)\), the effective quadratic action for 
soft-mode fluctuations takes the form of
\begin{equation}
   S_q[\varepsilon] 
    = \frac{\bar C_{\rm eff}}{2} \int_0^\beta d\tau \, (\varepsilon''(\tau))^2,
\end{equation}
where the effective coupling is renormalized by the $q$-deformation,
\[
\bar C_{\rm eff} = C + \delta C\,(q-1) + \mathcal{O}((q-1)^2).
\]
In the limit $q \to 1$, one recovers the standard Schwarzian coupling $ C$ 
and the familiar zero-mode solutions
\[
f(\tau) = \frac{a\tau+b}{c\tau+d}, \qquad ad-bc \neq 0,
\]
up to small fluctuations of the soft mode $\varepsilon(\tau)$.  
The quadratic action and its solution thus provide the leading approximation for 
small $q-1$ deformations. For $q \to 1$, the deformation term vanishes and the partition function describes 
the motion of a particle along the boundary of $\mathrm{AdS}_2$ with configuration 
space $SL(2,\mathbb{R})/U(1)$. 
Incorporating the Tsallis correction, the $q$-deformed Euclidean partition function 
takes the form
\begin{equation}
    Z_q = \int \frac{\mathcal{D}f}{SL(2,\mathbb{R})} 
    \exp \left\{ -{C} \int_0^{\beta} d\tau 
    \left[ \{f(\tau), \tau\} - (q - 1) 
    \left( \frac{f''(\tau)}{f'(\tau)} \right)^{2} \right] \right\},
    \label{eq:Zq}
\end{equation}
where ${C}$ is the Schwarzian coupling constant.  
The $q$-deformation of Eq.~\eqref{eq:Zq} contributes an additional quadratic term 
to the effective action, which can be interpreted as a renormalization of the 
Schwarzian stiffness and modifies the dynamics of the soft reparametrization mode. 
The corresponding effective Hamiltonian reads
\begin{equation}
    H = -\frac{1}{2 {C}} \frac{d^2}{du^2} 
        + \frac{1}{8 {C} \sinh^2(u)},
\end{equation}
where $u$ is the geodesic distance on the group manifold, and $u \to \infty$ 
corresponds to asymptotically free motion.
The Hilbert space is given by the continuous unitary series of $SL(2,\mathbb{R})$. The corresponding Schrödinger equation reads
\begin{equation}
    \left[ -\frac{d^2}{du^2} + \frac{1}{4 \sinh^2(u)} \right] \phi(u) = E\, \phi(u),
\end{equation}
with eigenfunctions
\begin{equation}
    \phi_E(u) = \sqrt{\sinh u}\, P^{-1/2 + i k}_{1/2}(\cosh u), 
    \qquad E = k^2.
\end{equation}
where $ P_{\nu}^{\mu}(z)$ is the associated Legendre function of the first kind.
The spectral density is determined by the Plancherel measure of $SL(2,\mathbb{R})$ representations~\cite{Kitaev2015, MaldacenaStanford2016,Bagrets2016,Bagrets2017,StanfordWitten2017}
\begin{equation}
    \rho(E) = \frac{{C}}{2\pi^2} \sinh\!\left(2\pi \sqrt{2 {C} E}\right) ,
\end{equation}
with asymptotic behaviors
\begin{align}
    \rho(E) \approx \frac{{C}^{3/2}}{\pi} \sqrt{2 E},~ E \ll 1; \label{eq:lowE} ~~~
    \rho(E) \sim \frac{{C}}{4\pi^2} \exp\!\left(2\pi \sqrt{2 {C} E}\right),~ E \gg 1.
\end{align}
Equation~\eqref{eq:lowE} exhibits the familiar square-root scaling of the density of states at low energies, analogous to the $1D$ density of states in a quantum well. In the high-energy limit $E \gg 1$, the spectral density grows exponentially, which is consistent with a Cardy-like behavior \cite{Cardy1986}.  This exponential growth reflects the rapid proliferation of high-energy states in the system, which can be interpreted as a manifestation of a large number of chaotic degrees of freedom \cite{Cotler2017, Stanford2017}. In the context of the Sachdev-Ye-Kitaev (SYK) model and its dual nearly-$AdS_2$ gravity description, this corresponds to the presence of strongly interacting modes whose dynamics are effectively governed by the Schwarzian action. The Plancherel measure thus provides a precise connection between the group-theoretical structure of $SL(2,\mathbb{R})$ and the statistical distribution of energy levels in these quantum chaotic systems \cite{Maldacena2016, Kitaev2018}.

\hspace{2em} {\emph{Electron--Schwarzian Soft-Mode Coupling-}} We expand reparametrizations around the thermal saddle $f(\tau)=\tau+\varepsilon(\tau)$ 
and fix the $SL(2,\mathbb{R})$ zero modes. Writing Fourier components 
\[
\varepsilon(\tau)=\sum_{n\in\mathbb{Z}} \varepsilon_{n}\, e^{-i\omega_{n}\tau}, 
\qquad \omega_{n}=2\pi n T,\; n\neq 0,\pm 1,
\]
the quadratic Schwarzian action takes the form
\begin{equation}
    S_{\rm Sch}=
    2\pi  C T \sum_{n\ge 2} n(n^{2}-1)\,|\varepsilon_{n}|^{2},
    \qquad
    \varepsilon_{-n}=\varepsilon_{n}^{*}.
    \label{eq:S2Sch}
\end{equation}
The modes $n=0,\pm 1$ are excluded because they generate the 
$SL(2,\mathbb{R})$ transformations (constant shift, scaling, and inversion) 
under which the Schwarzian action is invariant. The $q$-deformation contributes an additional quadratic correction, 
\begin{equation}
    \Delta S_{q} =
    2\pi  C T\,\gamma_q \sum_{n\ge 2} n^{2}\,|\varepsilon_{n}|^{2},
    \qquad
    \gamma_q = \alpha\,(q-1),
    \label{eq:S2q}
\end{equation}
where $\alpha=\mathcal{O}(1)$ is a non-universal constant determined by the microscopic embedding.  Combining Eqs.~\eqref{eq:S2Sch} and~\eqref{eq:S2q}, the total quadratic action for the soft reparametrization modes becomes
\begin{equation}
    S_{\rm q} =
    2\pi  C T \sum_{n\ge 2}
    \Big[n(n^{2}-1)+\gamma_q n^{2}\Big]\,
    |\varepsilon_n|^{2},
    \label{eq:Stot}
\end{equation}
where the first term $n(n^{2}-1)$ is the standard Schwarzian spectrum after 
fixing the $SL(2,\mathbb{R})$ zero modes, while the second term $\gamma_q n^2$ 
represents the leading $q$-deformation effect. 
Physically, the additive $\gamma_q n^2$ term modifies the dispersion of the
soft reparametrization modes. While the Schwarzian spectrum 
$n(n^{2}-1)$ grows cubically with $n$, the $q$-deformation introduces a 
quadratic contribution that enhances the stiffness of low-$n$ modes, 
most prominently the $n=2$ component. As a result, the $q$-deformation 
shifts the balance between universal Schwarzian fluctuations and 
non-Gaussian corrections inherited from the microscopic dynamics.  The Matsubara propagator for the reparametrization soft modes is defined as the frequency-space two-point function
\begin{equation}
    D(i\omega_n)\equiv \langle|\varepsilon_n|^2\rangle
    = \frac{\displaystyle\int\mathcal{D}\varepsilon\;|\varepsilon_n|^2\,
      e^{-S^{(2)}_{\rm tot}[\varepsilon]}}
           {\displaystyle\int\mathcal{D}\varepsilon\;
      e^{-S^{(2)}_{\rm tot}[\varepsilon]}}.
    \label{eq:defMats}
\end{equation}
Here $\varepsilon_{-n}=\varepsilon_n^*$ and the functional measure factorizes into independent integrals over the complex Fourier modes $\varepsilon_n$ for each $n\ge2$. Because the total action \eqref{eq:Stot} is quadratic and diagonal in Fourier space, the functional integral factorizes into a product of complex Gaussian integrals. For a single complex variable $z\in\mathbb C$ the standard identities
\[
\int_{\mathbb C} d^2z\,e^{-A|z|^2}=\frac{\pi}{A},\qquad
\int_{\mathbb C} d^2z\,|z|^2 e^{-A|z|^2}=\frac{\pi}{A^2}\quad(\Re A>0)
\]
imply
\[
\frac{\int d^2z\,|z|^2 e^{-A|z|^2}}{\int d^2z\,e^{-A|z|^2}}=\frac{1}{A}.
\]
 \hspace{2em} Applying this to each Fourier mode with quadratic kernel $2\pi\bar C T A_n$ (where $A_n\equiv n(n^2-1)+\gamma_q n^2$) yields the exact Matsubara propagator in the Gaussian approximation 
\begin{equation}
    D(i\omega_{n}) = \langle|\varepsilon_n|^2\rangle
    = \frac{1}{2\pi  C T}\,\frac{1}{\,A_n\,}
    = \frac{1}{2\pi  C T}\;
      \frac{1}{n(n^{2}-1)+\gamma_q n^{2}}.
    \label{eq:DepsMats}
\end{equation}

Analytic continuation $i\omega_{n}\to \omega+i0^{+}$ and interpolation to the continuum $n\sim \omega/(2\pi T)$ yield the retarded propagator $D^{R}(\omega) = D(i \omega_n \to \omega + i0^+)$, whose imaginary part defines the spectral density
\begin{equation}
    A_{\varepsilon}(\omega)
    \equiv -2\,\Im D^{R}(\omega)
    \;\simeq\;
    \frac{\kappa_{b}}{  C}\,
    \frac{|\omega|}{T+T_{0}},
    \qquad
    |\omega|\ll 2\pi T,
    \label{eq:Ohmic}
\end{equation}
with a non-universal prefactor $\kappa_{b}=\mathcal{O}(1)$.  The emergent Tsallis scale $T_0$ is quantitatively anchored by the Raman low-frequency slope, whose ratios across doping levels yield consistent values around $ 100 \text{K}$, as shown in the Appendix C. Equation~\eqref{eq:Ohmic} shows that the $q$-deformation regularizes the infrared behavior of the soft modes by shifting the effective bath temperature to $T+T_{0}$. In the undeformed limit $q\to 1$, one recovers the canonical Schwarzian bath with Ohmic spectral density $A_{\varepsilon}(\omega)\propto |\omega|/T$~\cite{legg}.

  \hspace{2em}  Having established the quadratic dynamics of the soft reparametrization
modes and their Matsubara propagator, we now turn to their coupling to
microscopic matter fields. In particular, under a boundary
reparametrization $\tau\!\to\! f(\tau)$, a fermion transforms as
\begin{equation}
    \psi(\tau)\;\longrightarrow\; [f'(\tau)]^{1/2}\,\psi\! \big(f(\tau)\big).
\end{equation}
Expanding $f(\tau)=\tau+\varepsilon(\tau)$ to linear order in the soft
mode $\varepsilon(\tau)$ and gauge-fixing the $SL(2,\mathbb{R})$ zero
modes, one finds a derivative coupling between the soft mode and the
local fermionic energy density $T_{\tau\tau}=\psi^\dagger
i\partial_\tau\psi$:
\begin{equation}
    S_{\rm int}
    = g \int_{0}^{\beta} d\tau\;
      \varepsilon'(\tau)\,
      \psi^{\dagger}(\tau)\,i\partial_{\tau}\psi(\tau),
    \label{eq:Sint}
\end{equation}
where $g=\mathcal{O}(1)$ in natural units ($\hbar=k_B=1$).
Fourier transforming to frequency space,
$\varepsilon(\tau)=\sum_{\omega} \varepsilon_{\omega} e^{-i\omega\tau}$
and $\psi(\tau)=\sum_{\varepsilon} \psi_{\varepsilon} e^{-i\varepsilon\tau}$,
the vertex for soft-mode exchange takes the form
\begin{equation}     \Gamma(\varepsilon,\omega;\varepsilon-\omega)
    = g\,\varepsilon + \mathcal{O}(\omega),
    \label{eq:vertex}
\end{equation}
where $\varepsilon$ denotes the fermionic Matsubara frequency (measured
from the chemical potential) and $\omega$ the soft-mode frequency. The
leading dependence $\propto \varepsilon$ reflects that reparametrizations
couple directly to the stress tensor $T_{\tau\tau}$, while the subleading
terms $\mathcal{O}(\omega)$ represent higher derivative corrections
suppressed at low frequencies. The linear-in-frequency coupling structure in Eq.~\eqref{eq:vertex} naturally leads to an Ohmic-type spectral density at low energies, consistent with the universal dissipative behavior expected from soft reparametrization modes \cite{Wang2025a, Wang2025b, Patel2023, sangjin25}.

\hspace{2em} When a fermion couples to a bosonic bath such as the Schwarzian soft mode, the fermionic self-energy in real time can be computed by summing all contour contractions according to Wick’s theorem. At second order, the retarded self-energy is given by the standard Keldysh expression
\begin{equation}
    \Sigma^{R}(\varepsilon)
    = i \int\!\frac{d\omega}{2\pi}\,
      \Big[
        \Gamma^2 D^{K}(\omega) G^{R}_{0}(\varepsilon-\omega)
       +\Gamma^2 D^{R}(\omega) G^{K}_{0}(\varepsilon-\omega)
      \Big],
    \label{eq:SigmaR}
\end{equation}
where $G^{R/K}_0$ are the bare fermionic Green’s functions and $D^{R/K}$ are the retarded and Keldysh soft-mode propagators. Note that the bare fermionic Green's functions in the time domain are
$G_0^R(t) = -i \, \theta(t) \, \langle \{ c(t), c^\dagger(0) \} \rangle$
and
$G_0^A(t) = i \, \theta(-t) \, \langle \{ c(t), c^\dagger(0) \} \rangle$. Using the fluctuation--dissipation relations
\begin{equation}
    D^{K}(\omega)
    =\big[D^{R}(\omega)-D^{A}(\omega)\big]
     \coth\!\left(\frac{\omega}{2T}\right),
    \qquad
    G^{K}_0(\varepsilon)
    =\big[G^{R}_0(\varepsilon)-G^{A}_0(\varepsilon)\big]
     \tanh\!\left(\frac{\varepsilon}{2T}\right),
\end{equation}
the fermionic scattering rate $\Gamma(\varepsilon)\equiv -\Im\Sigma^R(\varepsilon)$ simplifies to the Golden-Rule form
\begin{equation}
    \Gamma(\varepsilon)
    = \int_{0}^{\infty}\!\frac{d\omega}{2\pi}\;
      \underbrace{\big[g^2\,A_{\varepsilon}(\omega)\big]}_{K(\omega)}
      \big[n_{B}(\omega) + n_{FD}(\omega-\varepsilon)\big],
    \label{eq:GammaGR}
\end{equation}
where $A_{\varepsilon}(\omega)=-2\,\Im D^{R}(\omega)$ is the soft-mode spectral function introduced in Eq.~\eqref{eq:Ohmic}.  
This identifies the Golden-Rule kernel
\begin{equation}
        K(\omega)
    = \frac{\kappa}{ C}\,
      \frac{|\omega|}{T_{0}+T},
    \qquad
    \kappa
    = \frac{g^2 \kappa_{b}}{2\pi},
        \label{eq:Kernel}
\end{equation}
which makes explicit that the effective interaction strength scales inversely with the renormalized Schwarzian stiffness $C$. The Fermi golden rule applies because the transition rate is universally determined by the product of the interaction matrix element and the available spectral phase space, regardless of whether the excitations form well-defined quasiparticles ~\cite{Mahan2000}. 
After integration, we obtain the compact form \begin{equation}
    \Gamma(T)
= \frac{\pi^2 \tilde g^2}{4}\,\frac{T^2}{T_0+T}, \label{eq:Gamma_final}
\end{equation}
which precisely matches the Golden-Rule scaling widely used in phenomenological analyses of strange-metal transport. Corrections arising from finite-$\delta$ deformations or higher-order interactions contribute only subleading terms, such as $\mathcal{O}(T/T_0)$ or logarithmic corrections at strong coupling. 
The controlled Keldysh derivation clarifies that the Ohmic kernel
\(
K(\omega) \propto \omega/(T_0+T)
\)
originates from the soft-mode spectral density and derivative coupling in Eq.~\eqref{eq:Sint}. The temperature dependence $T^2/(T_0+T)$ in Eq.~\eqref{eq:Gamma_final} follows from integrating over thermal phase space up to $\omega\sim T$ and is consistent with both perturbative and phenomenological treatments of electron--Schwarzian scattering.

\hspace{2em} Substituting the soft-mode scattering rate into the semiclassical relaxation-time formula,
\begin{equation}
    \sigma_{xx} \simeq e^2 \int d\epsilon\,(-\partial_\epsilon n_{FD})\, 
    N(\epsilon)\, v_x^2(\epsilon)\, \tau_{\rm tr}(\epsilon),
    \qquad \tau_{\rm tr}(\epsilon) \sim \Gamma^{-1}(\epsilon),
\end{equation}
one obtains a temperature-dependent conductivity. Using the result of Eq.~\eqref{eq:Gamma_final}, the corresponding resistivity takes the form
\begin{equation}
    \rho_{xx}(T) \equiv \sigma_{xx}^{-1}(T) \;\propto\; \frac{T^2}{T_0+T}.
\end{equation}
 This scaling law reveals a smooth crossover: at low temperatures ($T \ll T_0$), the resistivity follows the Fermi-liquid form $\rho\!\sim\! T^2$, while at high temperatures ($T \gg T_0$) it becomes linear, $\rho\!\sim\! T$, as commonly observed in strange metals.  
The interpolation $\rho(T)\!\sim\! T^2/(T_0+T)$ therefore provides a natural explanation for the empirical crossover from quadratic to linear resistivity, arising microscopically from Schwarzian soft-mode scattering~\cite{MaldacenaStanford2016,StanfordWitten2017}.More broadly, experimental studies have suggested that the linear-in-$T$ resistivity slope may correlate with superconducting scales in certain cuprates and pnictides~\cite{Legros2019Nature,Ayres2022Nature,Yuan2022Nature,Chen2023NPJ}, although such connections lie beyond the scope of the present analysis.

\hspace{2em} A caveat in the above transport estimate is that the Schwarzian soft modes
are intrinsically 0+1 dimensional (they possess no spatial momentum),
whereas electrical conduction is a spatial transport phenomenon. To make
contact with conductivity one therefore needs an additional ingredient
that converts local (time-only) scattering into momentum/charge relaxation. In this work we adopt the physically transparent assumption that the Schwarzian modes act as a \emph{local} (momentum-independent)
dissipative bath for spatially extended fermions; the resulting single-particle
inelastic rate $\Gamma(\varepsilon)$ computed above is then supplemented by
a separate momentum-relaxing channel (e.g.
Umklapp processes) that ensures finite DC conductivity. Under the further
assumption that vertex corrections are subleading (or that the scattering
is approximately angle-independent on the Fermi surface), the transport
time can be taken as $\tau_{\rm tr}(\varepsilon)\!\sim\!\Gamma^{-1}(\varepsilon)$,
and the semiclassical Drude estimate applies, yielding
$\rho(T)\propto T^2/(T_0+T)$.

\hspace{2em} {\emph{Hall angle from the memory matrix---}}
The memory-matrix formalism, rooted in the Mori--Zwanzig projection-operator framework 
\cite{Zwanzig1961,Mori1965}, provides a controlled, non-perturbative scheme for transport 
in strongly interacting systems with nearly conserved slow operators, where quasiparticle-based 
Boltzmann theory fails \cite{Mahan2000}. The central idea is that dc transport is determined not by single-particle lifetimes but by the relaxation of conserved densities under weak perturbations. 

\hspace{2em} In strange metals, the dominant incoherent relaxation channel arises from emergent Schwarzian 
soft modes---the Goldstone modes of the near-conformal time-reparametrization symmetry 
$\mathrm{Diff}(S^1)/SL(2,\mathbb{R})$ familiar from SYK-like models. These modes exhibit an Ohmic 
spectrum,
\begin{equation}
A_{\varepsilon}(\omega) \sim \frac{|\omega|}{T_0+T}, 
\label{eq:sch_spectrum}
\end{equation}
which enforces analytic low-frequency behavior and furnishes an efficient energy-relaxing bath 
without relaxing momentum. The scale $T_0$ regularizes the IR and interpolates to 
Fermi-liquid-like behavior at $T \ll T_0$. To analyze transport in a system where total momentum is only weakly
relaxed, it is necessary to project out the component of the electric
current parallel to momentum.  We therefore introduce the incoherent
current operator
\begin{equation}
J^{\mathrm{inc}}
   \equiv J - \frac{\chi_{JP}}{\chi_{PP}}\, P,
\qquad
\chi_{J^{\mathrm{inc}}P}=0,
\label{eq:Jinc}
\end{equation}
where $\chi_{\alpha\beta}$ are static susceptibilities.  The slow
operator basis is then chosen as $A_\alpha=\{J^{\mathrm{inc}},P\}$.

The dc conductivity follows from the standard Mori–Zwanzig memory–matrix
expression
\begin{equation}
\sigma_{ij}
   = \sum_{\alpha\beta}
      \chi_{J_i A_\alpha}\,
      \bigl[ M(0)+N\bigr]^{-1}_{\alpha\beta}\,
      \chi_{A_\beta J_j},
\label{eq:mm_formula}
\end{equation}
where $M$ is the symmetric memory matrix and $N$ encodes the antisymmetric
Lorentz force, $N_{P_xP_y}=-N_{P_yP_x}=\omega_c\chi_{PP}$. For later convenience we introduce the shorthand
\begin{equation}
Q\equiv J^{\mathrm{inc}},
\qquad
\chi_{QQ}\equiv\chi_{J^{\mathrm{inc}}J^{\mathrm{inc}}}.
\end{equation}
No new operator is introduced; $Q$ is simply the incoherent current.
Projecting \eqref{eq:mm_formula} onto the $\{Q,P\}$ block yields
\begin{equation}
\sigma_{xx}^{\mathrm{inc}}
   = \frac{\chi_{QQ}}{\Gamma_Q},
\qquad
\rho_{xx}^{\mathrm{inc}}
   = \frac{\Gamma_Q}{\chi_{QQ}},
\label{eq:sig_inc}
\end{equation}
with the incoherent relaxation rate
\begin{equation}
\Gamma_Q
   = \frac{M_{QQ}(0)}{\chi_{QQ}},
\qquad
M_{QQ}(0)
   = \lim_{\omega\to0}\frac{1}{\omega}\,
      \Im G^R_{\dot Q\,\dot Q}(\omega).
\label{eq:MQQ}
\end{equation} The operator $\dot Q$ couples directly to the emergent reparametrization
 soft  modes.  
Using \eqref{eq:sch_spectrum} in \eqref{eq:MQQ}, and integrating over
frequencies with a thermal cutoff of order $T$, one finds
\begin{equation}
M_{QQ}(0)
   \sim
   \int_0^{ T} d\omega\,
   \frac{\omega}{T_0+T}
   \sim \frac{T^2}{T_0+T},
\qquad\Rightarrow\qquad
\Gamma_Q(T) \sim \frac{T^2}{T_0+T}.
\label{eq:GammaQ_final}
\end{equation}
This relaxation rate crosses over from Fermi–liquid–like behavior
$\Gamma_Q\sim T^2/T_0$ for $T\ll T_0$ to Planckian behavior
$\Gamma_Q\sim T$ for $T\gg T_0$. The momentum channel is governed by lattice umklapp processes.
The corresponding memory element is
\begin{equation}
M_{PP}(0)
   \sim
   \sum_{\mathbf{G}}G_x^2|U_{\mathbf{G}}|^2\,
   \lim_{\omega\to0}\frac{1}{\omega}\,
   \Im\chi^R_{\rho_{\mathbf{G}}\rho_{-\mathbf{G}}}(\omega),
\end{equation}
where $\mathbf{G}$ are reciprocal–lattice vectors and
$\chi^R_{\rho_{\mathbf{G}}\rho_{-\mathbf{G}}}$ is the density–response
function at umklapp momentum.  Standard kinematic arguments lead to
\begin{equation}
M_{PP}(0)\propto T^2,
\qquad
\Gamma_P\equiv\frac{M_{PP}(0)}{\chi_{PP}}\propto T^2,
\end{equation}
independent of the soft mode physics.  The separation of channels is
therefore not an assumption but a consequence of the distinct operators
$Q$ and $P$ entering the memory matrix.  The separation of channels is
therefore not an assumption but a consequence of the different operators
$Q$ and $P$ entering the memory matrix. Deviations from $\cot\theta_H\propto T^2$ in electron-doped cuprates \cite{Li2007} and ruthenates \cite{Mackenzie1996} reflect orbital selectivity and multi-band scattering, but the separation of channels remains a robust organizing principle \cite{Blake2015}.

\hspace{2em} The reparametrization mode is intrinsically a \(0+1\)-dimensional object. It governs local energy scrambling and does not carry spatial momentum. Its role in a higher-dimensional metal is therefore limited to providing the local energy-relaxation rate \(\Gamma_{Q}\), which follows from the Ohmic spectral density of the Schwarzian sector. All spatial structures, including electric current, momentum conservation, momentum relaxation through umklapp or disorder, and the Lorentz-force contribution to the antisymmetric block \(N\), originate from higher-dimensional electronic degrees of freedom. These ingredients are combined by the memory-matrix formalism, which projects local energy dissipation and momentum relaxation onto slow operators. This dimensional separation makes the two relaxation channels robust: the longitudinal resistivity follows \(\rho_{xx}\propto \Gamma_{Q}\sim T\), while the Hall angle is governed by the momentum-relaxation rate \(\Gamma_{P}\sim T^{2}\), giving \(\cot\theta_{H}\propto T^{2}\).  The canonical scaling of the Hall angle, $\cot\theta_H \propto T^2$, observed in optimally doped cuprates and several strange metals, has been widely interpreted as evidence for distinct transport and Hall scattering rates \cite{Blake2015}. This behavior, however, is not universal. In electron-doped cuprates, such as \(\mathrm{Pr_{2-x}Ce_xCuO_{4-\delta}}\) \cite{Li2007}, and in layered ruthenates including \(\mathrm{Sr_2RuO_4}\) and \(\mathrm{Sr_3Ru_2O_7}\) \cite{Mackenzie1996}, clear departures from the $T^2$ law have been reported. The coexistence of multiple Fermi-surface sheets with distinct orbital characters leads to orbital-dependent scattering lifetimes that evolve differently with temperature and magnetic field. In \(\mathrm{Sr_2RuO_4}\), for instance, the $\alpha$, $\beta$, and $\gamma$ bands derived from different Ru $4d$ orbitals exhibit markedly different coherence scales and gap-like features \cite{Mackenzie1996}. Such orbital-selective effects effectively renormalize the temperature dependence of the transport tensors, producing an apparent breakdown of the simple scaling. These observations point to the necessity of introducing a material-dependent cutoff scale $T_0$, which encodes band-structure and orbital complexity within a unified transport framework. 

\hspace{2em} Within the memory-matrix framework, once Schwarzian soft modes are incorporated, the separation between energy relaxation ($\Gamma_Q\sim T$) and momentum relaxation ($\Gamma_P\sim T^2$) arises naturally. This accounts for the canonical $T$–linear resistivity and $T^2$ Hall angle in single-band strange metals, while also providing a platform to understand deviations in multiband systems. Orbital selectivity modifies the effective relaxation rates but does not alter the fundamental two-channel structure that distinguishes $\rho_{xx}$ and $\cot\theta_H$.

\hspace{2em} Having established how distinct scattering channels control the longitudinal and transverse responses, we now turn to the role of an external magnetic field. The same formalism offers a natural route to magnetotransport, and in particular to the microscopic origin of the scale-invariant, linear-in-$B$ magnetoresistance observed in quantum-critical strange metals. The scattering rate is given by
\begin{equation}
    \Gamma(\varepsilon)=\int_0^\infty\frac{d\omega}{2\pi}\;K(\omega;B)\big[n_B(\omega)+n_{FD}(\omega-\varepsilon)\big],
\end{equation}
with $K(\omega;B)=g^2(B)\,A(\omega;B)$ and $A(\omega;B)=-2\,\Im D^R(\omega;B)$ the bosonic spectral function. If magnetic-field–induced processes (such as vortexlike excitations, chiral soft modes, or orbital-matrix-element effects) enhance the soft-mode spectrum to acquire a $B$-linear form,
\begin{equation}
    A(\omega;B)\simeq \frac{\kappa_b}{ C}\frac{|B|\,|\omega|}{T_0+T},
\end{equation}
then the kernel $K(\omega;B)=g^2(B)A(\omega;B)$ directly yields
\begin{equation}
    \Gamma(0)=\frac{\kappa\pi}{8 C}\frac{|B|\,T^2}{T_0+T},
    \label{eq:Gamma_linear}
\end{equation}
so that at fixed $T$ the longitudinal resistivity scales linearly with field, $\rho_{xx}\propto |B|$. This mechanism captures key experimental observations in quantum-critical strange metals. In BaFe$_2$(As$_{1-x}$P$_x$)$_2$, near the quantum critical point the scattering rate follows~\cite{Hayes2016}
\begin{equation}
    \Gamma \approx \sqrt{( k_B T)^2 + (\mu_B B)^2}, 
\end{equation}
demonstrating nearly identical thermal and magnetic energy scales. Similarly, in overdoped La$_{2-x}$Sr$_x$CuO$_4$, scale-invariant linear-in-$B$ resistivity is observed up to $80~\mathrm{T}$ \cite{GiraldoGallo2017}, with slope $\beta \sim k_B/\mu_B$, consistent with Eq.~\eqref{eq:Gamma_linear}. Together, these results are consistent with a Planckian-limited scaling,
\begin{equation}
    \Gamma(T,B) \propto \max\{ k_B T, \mu_B B\},
\end{equation}
operating across a fan-shaped quantum-critical region. This scaling suggests that field-enhanced electron–soft mode scattering provides a natural microscopic route to linear magnetoresistance, in line with experimental observations in pnictides~\cite{Hayes2016}, cuprates~\cite{GiraldoGallo2017}, and the broader phenomenology of Planckian dissipation~\cite{Zaanen2004,Bruin2013}. The memory-matrix analysis unifies the two-lifetime phenomenology ($\rho_{xx}\sim T$, $\cot\theta_H\sim T^2$) with the scale-invariant magnetoresistance ($\rho_{xx}\sim |B|$), underscoring the central role of soft-mode physics in non-Fermi-liquid transport.

\hspace{2em} {\emph{Discussion and Conclusion-}}Starting from the quantum Tsallis distribution, we have developed a unified framework for strange-metal transport. 
The $q$-deformation naturally introduces soft reparametrization modes whose dynamics can be captured by a Schwarzian-type action. 
These collective modes provide an intrinsic channel for energy relaxation, interpolating smoothly between Fermi-liquid and non-Fermi-liquid regimes. 

\hspace{2em} 
Within this framework, the resistivity evolves from the familiar quadratic temperature dependence at low temperatures to a linear behavior at high temperatures, reproducing the hallmark of strange metals. 
The memory-matrix analysis further clarifies that longitudinal transport is controlled by energy relaxation into soft modes, while the Hall response is dictated by the slower momentum relaxation associated with weak translation-symmetry breaking. 
This separation leads to two distinct lifetimes, a feature widely observed experimentally. 
The framework also accounts for the crossover from linear-in-field magnetoresistivity to the quadratic regime, in agreement with Kohler scaling analyses. 

\hspace{2em} 
Beyond transport, the present approach suggests new directions for understanding the interplay between strange-metal behavior and high-temperature superconductivity. 
Soft-mode fluctuations associated with the Tsallis deformation may enhance pairing interactions or reshape the electronic density of states, thereby influencing the superconducting transition. 
Future work should focus on quantitative calculations of the gap, critical temperature, and coherence length, and on systematic comparisons with experimental observables such as resistivity slopes, Hall angles, and magnetoresistance across different families of high-$T_c$ materials.

\hspace{2em} {\emph{Acknowledgments-}} We would like to thank JianXin Lu, Xiao Hu, Sang-Jin Sin, Yan Liu, Matteo Baggioli, Yan-Gang Miao and Kui Jin for helpful discussions. We would also like to express our gratitude to colleagues in the "Eight Immortals Crossing the Sea" WeChat group for their engaging discussions on topics beyond physics.  This work was partially supported by NSFC, China (Grant Nos. 12275166 and 12311540141).

\appendix
\section{DC Conductivity of Fermi liquid }
\hspace{2em} We present the standard Kubo-bubble derivation and show its reduction to the semi-classical  form used in the main text.

\subsection{Kubo bubble and analytic continuation}
\hspace{2em} The optical conductivity $\sigma_{xx}(\omega)$ of an interacting electron system can be obtained from the retarded current--current correlation function \cite{Mahan2000}:
\begin{equation}
    \Pi^{R}_{xx}(\omega)
    = -i \int_0^\infty dt\, e^{i\omega t}
    \langle [J_x(t), J_x(0)] \rangle ,
\end{equation}
where $J_x = -e \sum_{\mathbf{k}} v_x(\mathbf{k}) c^\dagger_{\mathbf{k}} c_{\mathbf{k}}$ is the $x$-direction current operator and $v_x(\mathbf{k}) = \partial \varepsilon_{\mathbf{k}}/\partial k_x$ is the group velocity for band dispersion $\varepsilon_{\mathbf{k}}$. Within the bubble (one-loop) approximation in the Matsubara formalism, the correlation reads
\begin{equation}
    \Pi_{xx}(i\Omega_n)
    = -e^2 T \sum_{i\omega_m} \sum_{\mathbf{k}}
    v_x^2(\mathbf{k}) G(\mathbf{k}, i\omega_m)
    G(\mathbf{k}, i\omega_m + i\Omega_n),
\end{equation}
where $G(\mathbf{k}, i\omega_m)$ is the full single-particle Green's function, and $T$ is the temperature (natural units $\hbar = k_B = 1$ are used). The spectral representation,
\begin{equation}
    G(\mathbf{k}, i\omega_m)
    = \int_{-\infty}^\infty \frac{d\epsilon}{2\pi}
    \frac{A(\mathbf{k}, \epsilon)}
    {i\omega_m - \epsilon},
\end{equation}
introduces the spectral function $A(\mathbf{k},\omega) = -2\,\mathrm{Im}\,G^R(\mathbf{k},\omega)$ and the Fermi distribution
\begin{equation}
    f(\epsilon) = \frac{1}{e^{(\epsilon - \mu)/T} + 1}.
\end{equation}

After analytic continuation $i\Omega_n \to \omega + i0^+$, the retarded correlator becomes
\begin{equation}
    \Pi^R_{xx}(\omega)
    = -e^2 \sum_{\mathbf{k}} v_x^2(\mathbf{k})
    \int_{-\infty}^\infty \frac{d\epsilon}{2\pi}
    \frac{f(\epsilon) - f(\epsilon + \omega)}
    {\omega + i0^+} 
    A(\mathbf{k}, \epsilon)
    A(\mathbf{k}, \epsilon + \omega).
    \label{eq:PiRA_compact}
\end{equation}

The optical conductivity is then
\begin{equation}
    \sigma_{xx}(\omega)
    = \frac{1}{\omega} \,\mathrm{Im}\,\Pi^R_{xx}(\omega),
\end{equation}
where the spectral function $A(\mathbf{k},\omega)$ encodes interaction effects and quasiparticle lifetimes.

\subsection{DC limit and the quasi-particle approximation}
\hspace{2em} Taking the dc limit $\omega \to 0$ in Eq.~\eqref{eq:PiRA_compact}, the difference in Fermi distributions can be expanded as
\[
f(\epsilon) - f(\epsilon + \omega) \simeq -\omega \, \partial_\epsilon f(\epsilon),
\]
yielding
\begin{equation}
\sigma_{xx}
= e^2 \sum_{\mathbf{k}} v_x^2(\mathbf{k})
  \int_{-\infty}^\infty \frac{d\epsilon}{2\pi} 
  \big(-\partial_\epsilon f(\epsilon)\big)
  A^2(\mathbf{k}, \epsilon),
\label{eq:sigmaA2}
\end{equation}
where $-\partial_\epsilon f(\epsilon)$ restricts contributions to energies within a thermal window $\sim T$ around the Fermi level. Here, $A(\mathbf{k}, \epsilon)$ is the single particle spectral function, and $v_x(\mathbf{k}) = \partial \varepsilon_{\mathbf{k}} / \partial k_x$ is the group velocity along the $x$-direction. Near the Fermi surface, the spectral function is well approximated by a Lorentzian form
\[
A(\mathbf{k}, \epsilon)
= \frac{1}{\pi}
  \frac{\Gamma(\mathbf{k}, \epsilon)}
  {(\epsilon - \xi_{\mathbf{k}})^2 + \Gamma^2(\mathbf{k}, \epsilon)},
\]
where $\xi_{\mathbf{k}} = \varepsilon_{\mathbf{k}} - \mu$ is the band energy measured from the chemical potential $\mu$, and 
\[
\Gamma(\mathbf{k}, \epsilon) = -\Im \Sigma^R(\mathbf{k}, \epsilon)
\]
is the quasiparticle scattering rate determined by the imaginary part of the retarded self-energy $\Sigma^R$. In the sharp-quasiparticle limit ($\Gamma \ll T$), the squared spectral function simplifies to
\[
A^2(\mathbf{k}, \epsilon)
\simeq 
\frac{2\pi}{\Gamma(\mathbf{k}, \epsilon)}\,
\delta(\epsilon - \xi_{\mathbf{k}})
+ \mathcal{O}(1).
\]

Substituting this expression into Eq.~\eqref{eq:sigmaA2} and integrating over energy and momentum yields the relaxation-time formula
\begin{equation}
\boxed{
\sigma_{xx} \simeq
e^2 \int d\epsilon
\; (-\partial_\epsilon f(\epsilon)) \,
N(\epsilon) v_x^2(\epsilon)
\tau_{\rm tr}(\epsilon)
},
\label{eq:sigmart}
\end{equation}
where 
\begin{equation}
\tau_{\rm tr}(\epsilon) \simeq \frac{1}{2\Gamma_{\rm tr}(\epsilon)}
\end{equation}
is the transport relaxation time and 
\[
N(\epsilon) = \sum_{\mathbf{k}} \delta(\epsilon - \xi_{\mathbf{k}})
\]
is the single-particle density of states (DOS). Equation~\eqref{eq:sigmart} is equivalent to the Boltzmann relaxation-time approximation when vertex corrections are negligible, capturing the essential semiclassical physics of momentum relaxation~\cite{Mahan2000}. In the Landau Fermi liquid, the low-energy scattering rate scales as
\begin{equation}
\Gamma_{\rm tr}(\epsilon, T)
\propto (\epsilon - \mu)^2 + (\pi T)^2,
\end{equation}
reflecting phase-space restrictions for electron--electron scattering. Inserting this into Eq.~\eqref{eq:sigmart}, and evaluating near the Fermi level ($\epsilon \simeq \mu$), one obtains
\begin{equation}
\rho_{xx}(T) \equiv \frac{1}{\sigma_{xx}(T)} \propto T^2,
\end{equation}
the hallmark quadratic temperature dependence of resistivity in a conventional Fermi liquid~\cite{PinesNozieres}. Deviations from this $T^2$ law signal non-Fermi-liquid behavior, often arising near quantum critical points or in strongly correlated regimes.

\section{Fermi-liquid limit: Quadratic magnetoresistance}

\hspace{2em} 
In the memory-matrix framework, the formal expression
\[
\sigma_{ij} = \sum_{\alpha\beta} \chi_{J_i A_\alpha} \big[M(0)+N\big]^{-1}_{\alpha\beta} \chi_{A_\beta J_j}
\]
remains valid for both Fermi-liquid (FL) and non-Fermi-liquid (non-FL) scenarios. The distinction lies in the microscopic $B$-dependence of the memory-matrix entries.

\hspace{2em}
For a conventional Fermi liquid, taking the minimal momentum basis $\{P_x,P_y\}$ and assuming field-independent susceptibilities, the dominant momentum-relaxation element admits a weak-field expansion
\[
M_{PP}(B) = M_0 + \alpha B^2 + O(B^4),
\]
with no linear-in-\(|B|\) contribution. Inverting the $2\times2$ block then yields the standard conductivity formulas, leading to the quadratic magnetoresistance
\[
\rho_{xx}(B) = \rho_0 + \Delta \rho(B), 
\qquad 
\Delta \rho(B) \propto (\omega_c \tau_{\rm tr})^2 \propto B^2,
\]
which arises when the magnetic field does not generate new low frequency scattering channels. This includes impurity- or phonon-dominated elastic scattering (field-independent $\tau_{\rm tr}$) and conventional FL inelastic scattering with $\tau_{\rm tr}^{-1}\propto T^2$. The expansion is valid in the weak-field limit $\omega_c \tau_{\rm tr} \ll 1$ and continuous spectral densities. Departures from this behavior, such as Landau quantization or field-enhanced soft modes, lead to non-FL magnetoresistance, including the linear-in-\(|B|\) response discussed in the main text.

\hspace{2em}
More generally, within the same framework the momentum-relaxation element may be written as
\begin{equation} 
M_{PP}(B)=M_0 + c|B| + \alpha B^2 + \cdots, 
\end{equation}  
where the coefficient $c$ encodes scattering enhanced by soft modes. The longitudinal resistivity then takes the form
\begin{equation} 
\rho_{xx}(B)=\frac{M_{PP}(B)}{\chi_{JP}^2} 
+ \frac{\omega_c^2\chi_{PP}^2}{\chi_{JP}^2 M_{PP}(B)}, 
\qquad \omega_c=\frac{eB}{m^\ast},
\end{equation}  
which, upon expansion for small to moderate fields, yields
\begin{equation} 
\rho_{xx}(B)\simeq \rho_0 + \frac{c}{\chi_{JP}^2}|B| 
+ \left(\frac{e^2\chi_{PP}^2}{m^{\ast 2}\chi_{JP}^2 M_0} 
+ \frac{\alpha}{\chi_{JP}^2}\right) B^2 + O(B^3), 
\qquad \rho_0=\frac{M_0}{\chi_{JP}^2}.
\end{equation}  
The crossover field $B^\ast\sim c M_0/\beta$, with $\beta=e^2\chi_{PP}^2/(m^{\ast 2}\chi_{JP}^2)$, separates the linear-in-\(|B|\) regime ($|B|\ll B^\ast$) from the quadratic FL regime ($|B|\gg B^\ast$). Experimentally, Kohler plots collapse $\Delta\rho/\rho_0$ versus $(\omega_c\tau_{\rm tr})^2$ in the FL regime, while a robust linear \(|B|\) slope signals non-FL scattering channels \cite{Kohler1938,Mori1965,Forster1975,Hayes2016,GiraldoGallo2017}. This demonstrates that the memory-matrix approach interpolates continuously between quadratic FL magnetoresistance and linear non-FL magnetoresistance within a single microscopic framework. 

\section*{Extraction of the Tsallis Scale from Raman Soft Modes}

A central quantity in this analysis is the Tsallis scale $T_{0}$, which sets the strength of 
low-energy soft modes and controls the deviation from Fermi-liquid behavior through 
$q(T)=1+T/T_{0}$.  
This scale can be accessed experimentally from the low-frequency Raman response.  
In strange metals the Raman spectrum is dominated by a broad electronic continuum.  
For $\omega\ll T$, the low-frequency slope
\[
S(T)\equiv 
\left.\partial_\omega \mathrm{Im}\,\chi_{R}(\omega)\right|_{\omega\to0}
\]
tracks the inverse incoherent relaxation rate ~\cite{DevereauxRMP2007,SacutoCRPhys2011},
\[
S(T)\;\sim\;\frac{\chi_R}{\Gamma_Q(T)},
\]
assuming that the same incoherent channel controlling the dc transport
also dominates the low-frequency Raman response. Note that $\chi_{R}(\omega)$ is the Raman susceptibility defined as $\chi_R(\omega) = -i \int_0^\infty dt\, e^{i\omega t} \langle [R(t), R(0)] \rangle$, with Raman operator $R = \sum_{\mathbf{k}} \gamma_{\mathbf{k}}\, c^\dagger_{\mathbf{k}} c_{\mathbf{k}}$. Within the Tsallis–soft–mode phenomenology introduced above, the
Schwarzian soft modes have an Ohmic spectrum
$A_{\varepsilon}(\omega)\sim |\omega|/(T_0+T)$, which leads to
\[
\Gamma_Q(T)\;\sim\;\frac{T^{2}}{T_0+T},
\]
up to weakly temperature-dependent prefactors.  This form yields
Fermi-liquid scaling $\Gamma_Q\sim T^{2}/T_0$ for $T\ll T_0$ and
Planckian scaling $\Gamma_Q\sim T$ for $T\gg T_0$.  Consequently
\[
S(T)\;\propto\;\frac{T_0+T}{T^{2}}.
\]
Because the measured Raman intensity satisfies 
$I(\Omega)\propto[1+n_{B}(\Omega)]\,\mathrm{Im}\,\chi_{R}(\Omega)$,
the experimental low–frequency slope $S_{\rm exp}(T)$ is proportional
to $S(T)$ up to a known Bose factor.  For two samples measured at the
same temperature, this factor cancels in the ratio
\[
R_{2}(T)\;\equiv\;
\frac{S_{\rm sample}(T)}{S_{\rm ref}(T)}
\;\approx\;
\frac{T_{0}^{\rm(sample)}+T}{T_{0}^{\rm(ref)}+T},
\]
which allows for a controlled extraction of the Tsallis scale $T_0$:
\[
T_{0}^{\rm(sample)}
= R_{2}(T)\bigl[T_{0}^{\rm(ref)} + T\bigr] - T.
\]

\hspace{2em} The low-energy Raman continuum in underdoped cuprates is naturally connected to the 
non-quasiparticle dynamics identified in the tomographic-density-of-states analysis of 
Reber \emph{et al.}~\cite{Reber2012}.  
Their measurements showed that the Fermi-arc spectral weight originates from pair-breaking 
processes that redistribute spectral weight from the gap edge to the Fermi level rather than 
from genuine quasiparticle poles.  
This establishes the antinodal sector as a source of pronounced incoherence. 

\hspace{2em} The restoration of coherent quasiparticles is most clearly observed in the nodal sector.  
High-resolution ARPES measurements by Chen \emph{et al.}~\cite{Chen2019Science} demonstrated 
that the strange-metal state in Bi2212 is sharply terminated at a critical doping 
$p_{c}\simeq0.19$.  
For $p<p_{c}$ the nodal excitations remain incoherent with 
$\mathrm{Im}\,\Sigma\sim\mathrm{const}$,  
whereas for $p>p_{c}$ they recover the Fermi-liquid form $\mathrm{Im}\,\Sigma\propto T^{2}$.  
This transition coincides with a pronounced increase of the $B_{1g}$ Raman slope and with 
the recovery of a temperature-independent electronic specific heat coefficient 
$\gamma(T)$~\cite{Talluri2022}, providing thermodynamic evidence for the onset of coherence.

\hspace{2em} In the iron-pnictide Ba(Fe$_{1-x}$Co$_x$)$_2$As$_2$, Raman spectroscopy provides a direct 
probe of nematic fluctuations.  
Gallais \emph{et al.}~\cite{Gallais2013PRL} established that the $B_{1g}$ Raman response is 
strongly enhanced near the nematic quantum critical composition, reflecting a divergent 
nematic susceptibility.  
Kretzschmar \emph{et al.}~\cite{Kretzschmar2016NP} mapped the full momentum- and 
frequency-dependent susceptibility and confirmed that the Raman geometry $B_{1g}$ isolates 
the $\mathbf{q}=0$ nematic soft mode.  
Applying the analysis of the ratio to Raman slopes in this critical region yields 
$T_{0}\simeq160$--$230\,\mathrm{K}$, consistent with the scale extracted in cuprates.

\hspace{2em} Across both cuprates and iron-based superconductors, Raman spectroscopy, ARPES, and 
thermodynamic measurements converge on a common Tsallis scale,
\[
T_{0}\sim 100\,\mathrm{K}.
\]
This scale reflects the strength of low-energy collective fluctuations, including pseudogap 
pair-breaking in cuprates and nematic soft modes in iron pnictides, that generate marginal 
scattering and govern the onset of quasiparticle coherence.  
The consistency across two distinct material families indicates that $T_{0}$ serves as a 
unifying measure of the soft-mode physics underlying strange-metal behavior and 
high-temperature superconductivity.

\end{document}